\documentclass[aps,prb,twocolumn,epsf,floatfix]{revtex4}
\usepackage{epsfig}
\usepackage{color}
\usepackage{graphicx}
\begin{document}

\preprint{cond-mat/07043863}

\title{Fano-Kondo effect through two-level system in triple quantum dots}
1
\author{Tetsufumi~Tanamoto and~Yoshifumi Nishi}

\affiliation{Advanced LSI laboratory, Corporate R\&D Center,
Toshiba Corporation,\\ 1, Komukai Toshiba-cho, Saiwai-ku,
Kawasaki 212-8582, Japan}

\begin{abstract}
We theoretically study the Fano-Kondo effect in a triple quantum dot (QD) system
where two QDs constitute a two-level system and the other QD works in a 
detector with electrodes. 
We found that the Fano dip is clearly modulated by strongly coupled QDs 
in a two-level system and a slow detector 
with no interacting QD. This setup suggests a new method of 
reading out qubit states. 
\end{abstract}
\maketitle

\section{Introduction}

Quantum dot (QD) systems have been providing opportunities to probe 
a wide variety of many-body effects of electronic transport 
properties in microelectronic structures.
The Kondo effect, one of the main characteristics of these systems, is 
a result of quantum correlation between localized spin in QD
and free electrons in electrodes\cite{Tarucha}.
Recently, the Fano effect, which appears as a result of quantum interference 
between a discrete single energy level and a major electronic system,
has also attracted the interests of many microelectronics researches.
Typically, a dip structure can be observed 
in conductance plotted as a function of an energy level of a side QD\cite{Gores,Sato,Kang,Aligia}. 

A T-shaped QD system consists of two QDs, in which 
the first QD is connected to a source and a drain and 
the second QD is set beside the first QD.
This system is well suited to investigation of both the Fano effect 
and Kondo effect (Fano-Kondo effect) and 
has been theoretically analyzed by 
many authors\cite{Wu,Guclu,Tanaka}.
There are a variety of parameter settings in a T-shaped QD system. 
Wu {\it et al.} assume that there is an infinite on-site
Coulomb interaction in the first QD 
(hereafter we call it `detector QD'), and no interaction in the 
second side-QD\cite{Wu}. On the other hand, 
G\"{u}cl\"{u} {\it et al.} assumes an infinite on-site 
Coulomb interaction in the second side-QD and no interaction in the 
detector-QD\cite{Guclu}.
In the former case, Kondo peak is dominant, whereas in the latter case, 
Fano dip structure is dominant in the transport properties of such a system.
Tanaka {\it et al.} assume infinite on-site Coulomb interactions in  
both the detector-QD and the side-QD, showing that Fano dip 
structure appears in the middle of Kondo resonance peak\cite{Tanaka}.

Here we theoretically investigate the Fano-Kondo effect in a triple QD system, 
depicted in Fig. 1.
When two QDs are coupled, the energy levels mix, and 
bonding and anti-bonding states are formed.
Thus, it is expected that the current that flows through the detector-QD 
reflects these electronic states when coupling ratio $t_C/t_d$ is sufficiently large, 
and the Fano dip would thus be modulated due to this two-level system.
We can also regard this setup as an extra impurity connected to a T-shaped QD system, 
or a charge qubit closely attached to a QD detector.
Coupled QDs can be used as a charge qubit\cite{tana0,Goan,Korotkov,tanahu,Gilad}.
Charge qubits are usually capacitively coupled to 
detectors that are switched on only after quantum computation is finished. 
Results of quantum computation are readout through the field-effect 
(Coulomb interaction) depending on the positions of electrons 
in the charge qubit\cite{tana0,Goan,Korotkov,tanahu,Gilad}.
If transfer of an electron between a charge qubit and a detector is not 
prohibited ($t_d \neq 0$) as in the present setup, 
detector current is expected to be changed by the Fano-Kondo effect.
This is an alternative method for reading out the quantum state of the charge 
qubit, although this method might change the quantum 
state of qubit more than the previous one, because detector 
electrons go through the charge qubit directly.
To operate a coupled QD system as a charge qubit, 
we have to maintain the electronic state of the two QDs 
near the degeneracy point of charging energy, or 
hold the total number of excess electrons in the charge 
qubit to one. These requirements lead to stronger constraints on this system. 
Here, as a first step,  we  consider only an effect of a two-level system 
without discussing qubit behaviors.
\begin{figure}
\includegraphics[width=6cm]{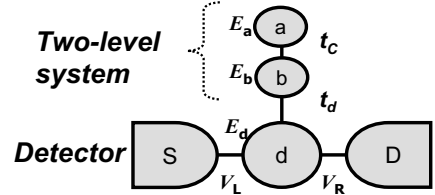}
\caption{Schematic plot of QD system. 
QDs $a$ and $b$ constitute a two-level system that is  
coupled to QD $d$ only which is connected to 
the electrodes. We consider two cases with differing 
on the strength of on-site Coulomb interaction 
in QD $d$. In case I, there is no on-site 
Coulomb interaction in QD $d$. In case II,
there is infinite on-site Coulomb interaction 
in QD $d$.
}
\label{System}
\end{figure}

A. W. Rushforth {\it et al.} experimentally investigated the Fano-Kondo effect 
when there are two side QDs near the electronic transport region\cite{Rushforth}. 
They showed that Fano dip is modulated by the path of electrons through two side QDs.
Compared with a simple T-shaped QD system, there are more 
parameters in this system due to the existence of extra QD.
Correspondingly, the Fano-Kondo effect in this system 
is expected to show more variety of dip behaviors than 
that in the T-shaped QDs as shown in Ref.\cite{Rushforth,Sasaki}.
To make the problem transparent without loss of generality,
we assume that there is a single energy level in each QD and 
that the two energy levels of the QD $a$ and QD $b$ coincide 
and correspond to gate voltages that are applied to those QDs.
For convenience, hereafter we call QD $a$ and $b$ a ``two-level system". 
We mainly investigate the change of conductance as a function 
of the gate bias. 

We use slave boson mean field theory (SBMFT) with the help 
of nonequilibrium Keldysh Green's functions. The formulation 
of SBMFT is very useful and a good starting point 
to study the transport properties 
of a strongly correlated QD system. Thus, this method is widely used, 
although it is only usable at a lower temperature 
($T$) region than the Kondo temperature $T_K$\cite{Newns,Lopez},
that is $T<T_K$.
Here, we investigate two cases differing in the amount of Coulomb 
interaction in QD $d$. In case I, there is no Coulomb 
interaction ($U_d=0$) and in case II, there is an infinite Coulomb
interaction ($U_d=\infty$). The case I experimentally corresponds to 
a large QD $d$ and the case II corresponds to a small QD $d$.
It is expected that 
we can see clear Fano effect in case I, and a 
Kondo resonance peak strongly changes Fano dip in 
case II.
Actually, it will be shown that, in order to see the effects of bonding-antibonding 
states in the detector current, coupling between the two side QDs 
should be strong and the detector should be sufficiently slow 
only in case I.
 
\section{Formulation}
The Hamiltonian for the case II is different from that for the case I 
in that the former has an additional constraint on QD $d$.
The mean field Hamiltonian for the case II is described in terms 
of slave bosons $b_{\alpha_1}$ 
$(\alpha_1=a,b,d)$ as:  
\begin{widetext}
\begin{eqnarray}
\lefteqn{
H^{\rm (II)}=\sum_{\alpha=L,R}\sum_{k_\alpha,s}
E_{k_\alpha}c_{k_\alpha s}^\dagger c_{k_\alpha s}
+\sum_{\alpha_1=a,b,d}\sum_{s}
E_{\alpha_1}f_{\alpha_1s}^\dagger f_{\alpha_1s}
+\sum_{\alpha_1=a,b,d}
\lambda_{\alpha_1} \left(\sum_s f_{\alpha_1s}^\dagger f_{\alpha_1s}
+b_{\alpha_1}^\dagger b_{\alpha_1}-1\right)
}\nonumber \\
\!&+&\!\frac{t_C}{N}\sum_{s}(
 f_{as}^\dagger b_ab_b^\dagger f_{bs}
+f_{bs}^\dagger b_bb_a^\dagger f_{as})
+\frac{t_d}{N}\sum_{s}(
 f_{ds}^\dagger b_db_b^\dagger f_{bs}
+f_{bs}^\dagger b_bb_d^\dagger f_{ds})
+\sum_{\alpha=L, R}\frac{V_\alpha}{\sqrt{N}} \sum_{k_\alpha,s}
(c_{k_\alpha s}^\dagger b_d^\dagger f_{ds}
+f_{ds}^\dagger b_d c_{k_\alpha s})
\label{Hamiltonian}
\end{eqnarray}
\end{widetext}
where $E_{k_\alpha}$ is the energy level for source ($\alpha=L$) and drain ($\alpha=R$) 
electrodes. $E_{a}$, $E_{b}$ and $E_{d}$ are energy levels for the triple QDs, respectively.
$t_C$, $t_d$ and $V_\alpha$ are the tunneling coupling strength 
between QD $a$ and QD $b$, that between QD $b$ and QD $d$, and 
that between QD $d$ and electrodes, respectively.
$c_{k_\alpha s}$ and $f_{\alpha_1s}$ are annihilation operators of the electrodes, 
and of the triple QDs $(\alpha_1=a,b,d)$, respectively.
$s$ is spin degree of freedom with spin degeneracy $N$; here we apply $N=2$.
$\lambda_{\alpha_1}$ is a Lagrange multiplier.
We take $z_{\alpha_1} \equiv b_{\alpha_1}^\dagger b_{\alpha_1}/2$ and 
$\tilde{E}_{\alpha_1} \equiv E_{\alpha_1}+\lambda_{\alpha_1}$
as mean field parameters.
The Hamiltonian for case I is similar to $H^{\rm (II)}$ except 
that $\lambda_d=0$ and $b_d=1$ in Eq.(\ref{Hamiltonian}).

By applying a conventional procedure\cite{Newns,Lopez} 
to this Hamiltonian, we derive Green's functions for QDs, for example, 
$
G_{dd}^r(\omega) =D_{ab}/B_{00}
$
{\it etc.}, 
where 
$D_{ab}\equiv (\omega-\tilde{E}_a)(\omega-\tilde{E}_b)
-\tilde{t}_C^2$, 
$B_{00} \equiv {D_{ab}B_r-(\omega-\tilde{E}_a)|\tilde{t}_d|^2}$
and $B_r\equiv \omega-\tilde{E}_d+i\Gamma$ with 
$\tilde{t}_C=t_C b_a b_b^\dagger/N$ and
$\tilde{t}_d=t_d b_d b_b^\dagger/N$ (see Appendix).
Here, $\Gamma_\alpha\equiv 2\pi \rho_\alpha (E_F)|V_\alpha |^2$ is the tunneling rate 
between $\alpha$ electrode and QD $d$ with a density of states (DOS), $\rho_\alpha (E_F)$,
for each electrode at Fermi energy $E_F$.
$\Gamma\equiv(\Gamma_L+\Gamma_R)/2$ and 
we assume $\Gamma_L=\Gamma_R$.

Four self-consistent equations for case I are given.
\begin{eqnarray}
& &\tilde{t}_C \sum_{s}
\langle f_{bs}^\dagger  f_{as} \rangle 
\!+\!\lambda_a |b_a|^2 =0, 
\label {s_eq1_1}\\
& &\tilde{t}_C \sum_{s}
\langle f_{as}^\dagger f_{bs} \rangle
+\tilde{t}_d \sum_{s}
\langle f_{ds}^\dagger f_{bs} \rangle
\!+\!\lambda_b |b_b|^2 =0, 
\label {s_eq1_2}\\
& & \sum_s \langle f_{\alpha_1 s}^\dagger f_{\alpha_1 s}\rangle 
+|b_{\alpha_1 }|^2=1, 
\ \ (\alpha_1=a,b),
\label {s_eq1_3}
\end{eqnarray}
For case II, we have to add two more equations regarding constraints on QD $d$
with $\tilde{V}_\alpha=V_\alpha b_d/\sqrt{N}$:
\begin{eqnarray}
\tilde{t}_d^* \sum_{s}
\langle f_{bs}^\dagger f_{ds}\rangle
\!\!\!&+& \! \! \! \! \!\!\!\sum_{\alpha=L,R} \sum_{k_\alpha,s}\tilde{V}_\alpha^* 
\langle c_{k_\alpha s}^\dagger f_{ds} \rangle
\!+\!\lambda_d |b_d|^2\!=\!0 
\label {s_eq2_1}\\
\sum_s \langle f_{ds}^\dagger f_{ds}\rangle &+& |b_{d}|^2=1.
\label{s_eq2_2}
\end{eqnarray}

To see the Fano-Kondo effect, we calculate conductance $G\equiv dI_L/dV$ 
at zero bias $V=0$. Source current $I_L$ is calculated from Green's functions as
$
I_L=\frac{e}{\hbar}\int \frac{d\omega}{\pi}
\frac{|D_{ab}|^2z_d^2}{B_{00}} \Gamma_R\Gamma_L (f_L(\omega)-f_R(\omega))
$ with $f_L (\omega)\equiv [\exp ((\omega+eV)/T)+1]^{-1}$ and
$f_R (\omega)\equiv [\exp (\omega/T)+1]^{-1}$.
Then, we obtain a conductance formula:
\begin{equation}
G=-\frac{e}{\hbar}\int \frac{d\omega}{\pi}
\frac{|D_{ab}|^2z_d^2}{B_{00}} \Gamma_R\Gamma_L \frac{\partial f_L(\omega) }{\partial\omega}
\label{cond}
\end{equation}
The ratio $t_C/t_d$ compares the internal coupling strength in a two-level system
with that between the two-level system and the detector.
Here, we regard the case where $t_C/t_d=5$ as  
a strongly coupled two-level system and the case where $t_C/t_d=1$
as a weakly coupled two-level system.
Because $\Gamma_L\Gamma_R/(\Gamma_L+\Gamma_R)$ ($=\Gamma$ ; we 
assume $\Gamma_L=\Gamma_R$) can be approximately estimated 
as a tunneling rate through the detector,
 if $\Gamma /t_d$ is sufficiently large, the electron 
that flows through QD $d$ is so fast that it cannot detect the oscillation of 
an electron in the coupled QDs $a$ and $b$.
In contrast, if $\Gamma /t_d$ is small, the electron 
that flows through QD $d$ can observe the oscillation between bonding and antibonding 
states.
Therefore, we call a detector with large $\Gamma/t_d=2$ a fast detector,
and one with smaller $\Gamma/t_d=0.4$ a slow detector.
$T_K$ is estimated as 
$T_K\sim D e^{-\pi |\tilde{E}_d-E_F|/\Gamma}$ 
$\sim 1.6 t_d$ ($D$ is a bandwidth), when we assume that $D= 20 t_d$, $ |E_d| <0.4t_d$, $\Gamma>0.4t_d$ 
and $E_F=0$.

\section{Numerical calculations}
Here, we numerically investigate the transport properties 
of our triple QD system for case I ($U_d=0$) and case II 
($U_d=\infty$). 

\begin{figure}
\includegraphics[width=7.5cm]{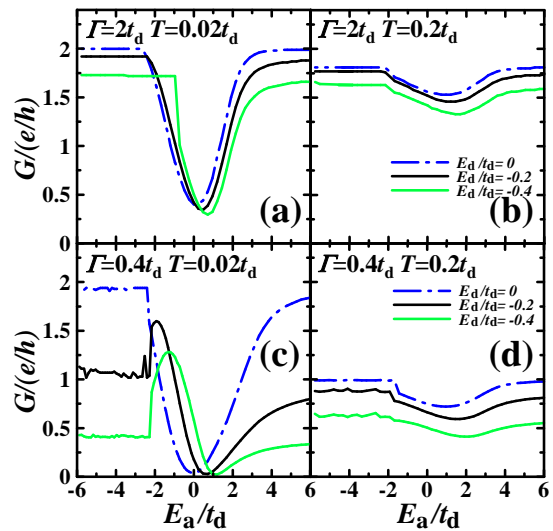}
\caption{Conductance $G$ as a function of an energy level
of the two-level system $E_a (=E_b)$ with 
a strong coupling ($t_C/t_d=5$) in case I ($U_d=0$).
(a) $T/t_d=0.02$ and (b) $T/t_d=0.2$ for a fast detector 
($\Gamma/t_d=2$).
(c) $T/t_d=0.02$ and (d)$T/t_d=0.2$ for a slow detector 
($\Gamma/t_d=0.4$).
Comparison of (a) with (c) shows that the dip structure 
is modulated in a slow detector when energy level $E_d$ is changed.
 $E_F=0$. Numerical results are obtained in $10^{-4}$ relative error range.}
\label{Strong}
\end{figure}
\begin{figure}
\includegraphics[width=7cm]{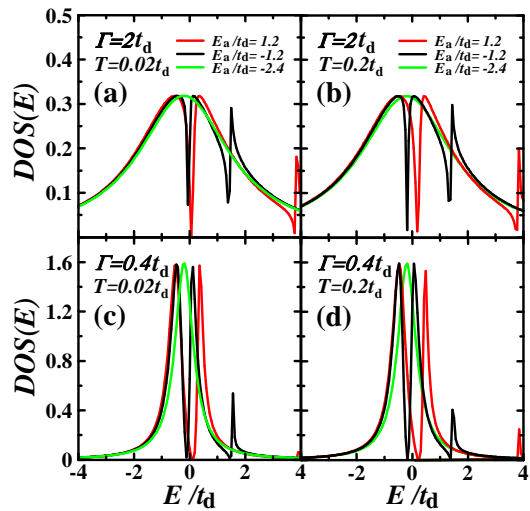}
\caption{Density of states Eq.(\ref{dos}) 
for a strong coupling ($t_C/t_d=5$) in case I ($U_d=0$).
(a) $T/t_d=0.02$ and (b) $T/t_d=0.2$ for a fast detector 
($\Gamma/t_d=2$).
(c) $T/t_d=0.02$ and (d)$T/t_d=0.2$ for a slow detector 
($\Gamma/t_d=0.4$).
$E_d=-0.2t_d$.$E_F=0$.}
\label{DosDos}
\end{figure}

\begin{figure}
\includegraphics[width=7cm]{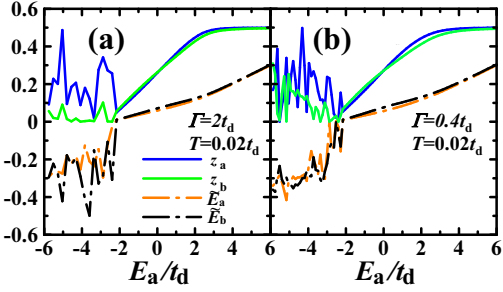}
\caption{Mean field values as a function of $E_a (=E_b)$: solutions 
of the self-consistent equations in case I.
(a) fast detector. (b) slow detector. $T=0.02t_d$. $E_d=-0.2t_d$. $t_d=0.05$. $E_F=0$.}
\label{MeanField}
\end{figure}

\begin{figure}
\includegraphics[width=7.5cm]{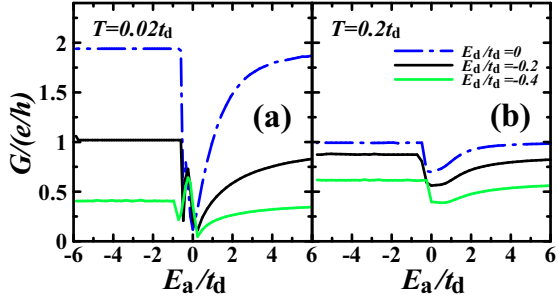}
\caption{Conductance $G$ as a function of $E_a (=E_b)$ for 
a weak coupling ($t_C/t_d=1$) in case I ($U_d=0$).
(a) $T/t_d=0.02$ and (b)$T/t_d=0.2$ for a slow detector 
($\Gamma/t_d=0.4$). 
Figure (a) shows that the dip structure 
is modulated for a slow detector when energy level $E_d$ is changed.
$E_F=0$. }
\label{WeakSlow}
\end{figure}
\subsection{Case I: $U_d=0$}
Figure~\ref{Strong} shows a conductance $G$ for strongly coupled
two-level system ($t_C/t_d=5$) in case I.
For a fast detector with $\Gamma/t_d =2$ (Fig.~\ref{Strong}(a) and (b)),
clear Fano dip can be seen in both low temperature
($T/t_d=0.02$) and high temperature ($T/t_d=0.2$).
However, we cannot see the effect of the change caused by the bonding-antibonding 
state
, even if we change $E_d$.
When $\Gamma$ is reduced as shown in Fig.\ref{Strong} (c), 
we can see that the Fano dip is changed from a simple dip form to 
an asymmetric form at low temperature.
The modulation of the structure in the $E_d<0$ region is considered to be 
an effect of the bonding-antibonding oscillation in the two-level system.
These behaviors can be understood from DOS, which 
is derived from $\rho_d (\omega)=-{\rm Im}G_{dd}^r (\omega)/\pi$ :
\begin{equation}
\rho_d(\omega)=\frac{(\Gamma/\pi)D_{ab}^2}
{\tilde{t}_d^4(\omega-\tilde{E}_a)^2
-2\tilde{t}_d^2(\omega-\tilde{E}_d)(\omega-\tilde{E}_a)D_{ab}
+D_{ab}^2 |B_r|^2}.
\label{dos} 
\end{equation}
Instead of a single dip in the numerator in a T-shaped QDs\cite{Wu},
$D_{ab}$ in the numerator of Eq.(\ref{dos}) indicates double dips, 
that comes from bonding-antibonding state of side-QDs. 
Figure~\ref{DosDos} shows four DOSs, which are displayed corresponding to Fig.\ref{Strong}.
We can see that there is a single peak for $E_a/t_d=-2.4$, 
closely distributed three peaks for $E_a/t_d=-1.2$, and 
separated three peaks for $E_a/t_d=1.2$. 
By comparing Figs.~\ref{DosDos}(a) and (c) with Figs.~\ref{DosDos}(b) and (d), respectively, 
we can see that temperature dependence on the shape of DOS is weak. Temperature 
slightly shifts the peak position.
In contrast, $\Gamma$ dependence is large, similar to a Breit-Wigner function.  
The DOS of the slow detector (Fig. \ref{DosDos} (c)(d)) is sharper than 
that of the fast detector (Fig. \ref{DosDos} (a)(b)). 
$\Gamma$ mainly determines the width and the maximum value similar to 
T-shaped QD system.
Because Eq.(\ref{cond}) is expressed as 
$
G=-(2e/\hbar)\int d\omega
\rho_d (\omega) z_d \Gamma_R\Gamma_L /(\Gamma_R+\Gamma_L)
\partial f_L(\omega) /\partial\omega
$, 
conductance reflects an energy region $\pm T$ around $E_F$ in DOS. 
From $\partial f_L(\omega) /\partial\omega\rightarrow \delta (\omega )$
when $T\rightarrow 0$, 
at low temperature, 
conductance can detect the change of closely distributed three peaks 
in the DOS if those peaks are sufficiently sharp such as a slow detector. 
On the other hand, the present results show that the peak structure 
for a fast detector is not sufficiently sharp 
to detect the three peaks. 
At higher temperature, conductance integrates  
peak structures in DOS, resulting in the single-dip structures 
shown in Fig.\ref{Strong} (b) and (d).
These are reasons of higher sensitivity of the slow detector at low temperature.

Figure ~\ref{MeanField} also shows the difference 
between a slow detector and a fast detector. 
Oscillations of mean field values for a fast detector 
(Fig.\ref{MeanField} (a) ) are weaker than those 
for a slow detector (Fig.\ref{MeanField} (b)) at $E_a<0$.

Figure \ref{WeakSlow} shows the case of weak coupling ($t_C/t_d=1$) 
in a slow detector. 
Fano dip modulation exists at lower temperature,
but is smaller than that of Figure~\ref{Strong}(c).
At the higher temperature region of $T=0.2t_d$, we can observe a  
small and calm dip structure for both fast and slow detectors.

\subsection{Case II: $U_d=\infty$}
\begin{figure}
\includegraphics[width=7.5cm]{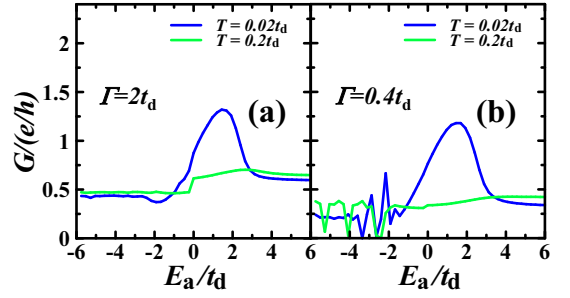}
\caption{Conductance $G$ as a function of an energy level
of the two-level system $E_a (=E_b)$ for 
a strong coupling ($t_C/t_d=5$) in case II ($U_d=\infty$).
(a) $\Gamma/t_d=2$ and (b)$\Gamma/t_d=0.4$. $E_d=E_F=0$. }
\label{3Dot}
\end{figure}

In this case, full self-consistent
equations are applied so that the total number of 
electrons in QD $d$ is less than one, and 
we have to solve the six self-consistent 
equations from Eq.(\ref{s_eq1_1}) to Eq.(\ref{s_eq2_2}). 
Without QD $a$ and QD $b$, 
a Kondo peak is expected. In the case of T-shaped QDs, 
the Fano dip appears in the middle of 
Kondo peak conductance\cite{Tanaka}.  
In the present case,
the two energy levels in the side-QDs couple with the Kondo peak and 
produces the complicated conductance structure shown 
in Fig.\ref{3Dot}. Here, overall structure of $G$ 
does not change even in the region of $E_d<0$.

\section{Discussion}
As noted in the introduction, the present model can be 
applied to several situations. 
If we look at QD $a$ as an extra trap site for the T-shaped QD system
composed of QD $b$ and QD $d$ with both electrodes, 
the present calculation shows that 
a simple dip structure is most likely to be observed, and 
when the extra trap site is strongly coupled to 
the side QD $b$ and the energy level of QD $d$ is sufficiently low, 
the modulation of Fano dip could be observed.
Experimentally, this would happen because it is not easy to 
exclude all trap sites from the QD system due to the difficulty of nanofabrication\cite{Gores}.
The present calculations also indicate that when there are many additional 
trap sites, the Fano dip structure would be modulated further. 

To construct a charge qubit\cite{tana0,tanahu} in the present case, 
the total number of electrons in QD $a$ and QD $b$ should be one. 
However, in the present setup, we only apply less than one electron in each QD. 
Thus, we need more restriction to hold one electron through 
the QD $a$ and QD $b$, for example, by introducing an exchange interaction 
such as that discussed in 
Ref.\cite{Lopez}. 

Although SBMFT is believed to be valid in previous works for 
one or two QDs, it has not yet been proved that this approximation
is valid for many QD systems in the context of the Fano-Kondo effect. 
The effect of fluctuations around mean fields should be
investigated in the near future.
%

\section{Conclusion}
We have studied the transport properties of a triple QD system, 
in particular, 
emphasizing the Fano-Kondo effect through two-level system.
We have used slave-boson mean field theory to describe 
quantum interference between electrons in discrete energy levels
and free electrons in electrodes.
We found that the Fano dip structure is modulated from simple dip to
clear asymmetric form,
when the coupling strength between the QDs in two-level system is large, 
using a slow detector without on-site Coulomb interaction in the detector QD.
We also found that the Fano dip is strongly changed 
by Kondo resonance peak when there is an infinite Coulomb interaction in the detector QD.

\acknowledgements
We are grateful to A. Nishiyama, J. Koga, S. Fujita,  
R. Ohba and M. Eto for valuable discussion.

\appendix
\section{Green's functions}
Green's functions are obtained based 
on equations of motion for the time ordered Green's functions.
Fourier-transferred diagonal Green's functions for QDs are given as
$
G_{aa}^r(\omega)=
[(\omega-\tilde{E}_b)B_r -|\tilde{t}_d|^2]/B_{00}
$ 
and 
$
G_{bb}^r(\omega)=
[(\omega-\tilde{E}_a)B_r]/B_{00}
$ other than $G_{dd}^r(\omega)$ expressed in the main text.
We also have off-diagonal Green's function by using 
relations 
$
(\omega -\tilde{E}_a ) G_{ab}(\omega)=\tilde{t}_C G_{bb}(\omega)
$
and 
$
(\omega -\tilde{E}_b -|\tilde{t}_d|^2/(\omega -\tilde{E}_a)  ) G_{ab}(\omega)
=\tilde{t}_d G_{dd}(\omega)
$.
$G_{dk_\alpha}(\omega)$ and $G_{k_\alpha k_\alpha'}(\omega)$($\alpha=L,R$) 
are obtained by 
$
\{ \omega-\tilde{E}_d -\Sigma_b 
-|\tilde{t}_d|^2/[\omega -\tilde{E}_b -|\tilde{t}_d|^2/(\omega -\tilde{E}_a)  ]
\}G_{dk_\alpha}(\omega)=V_\alpha /(\omega-E_{k_\alpha})
$ 
and 
$
(\omega-E_{k_\alpha}) G_{k_\alpha k_\alpha'}(\omega)=\delta_{k_\alpha k_\alpha'}
+V_\alpha G_{dk_\alpha}(\omega)
$, with 
$
\Sigma_b \equiv \sum_{\alpha}\sum_{k_\alpha} |V_\alpha|^2/(\omega-E_{k_\alpha})
$.
After applying analytic continuation rules, we obtain lesser Green's functions
in the current formula in the main text.

\end{document}